\documentclass[%
 reprint,
 amsmath,amssymb,
 aps,
]{revtex4-2}

\usepackage{graphicx}% Include figure files
\usepackage{dcolumn}% Align table columns on decimal point
\usepackage{bm}% bold math
\usepackage{physics} % Ju added it
\usepackage[normalem]{ulem} % Ju added it
\usepackage{xcolor} % Ju added it

\begin{document}

\preprint{APS/123-QED}

\title{Quantitative study of enantiomer-specific state transfer}% Force line breaks with \\

\author{JuHyeon Lee,\textit{$^{1}$} Johannes Bischoff,\textit{$^{1}$} A. O. Hernandez-Castillo\textit{$^{1}$}, Boris Sartakov,\textit{$^{1,2}$} Gerard Meijer,\textit{$^{1}$} and Sandra Eibenberger-Arias\textit{$^{1}$}}
\email{eibenberger@fhi-berlin.mpg.de}

\affiliation{%
\textit{$^{1}$}Fritz-Haber-Institut der Max-Planck-Gesellschaft, Faradayweg 4-6, 14195 Berlin, Germany\\
\textit{$^{2}$}Prokhorov General Physics Institute, Russian Academy of Sciences, Vavilovstreet 38, 119991 Moscow, Russia\\
}%

\date{\today}% It is always \today, today,

\begin{abstract}
We here report on a quantitative study of Enantiomer-Specific State Transfer (ESST), performed in a pulsed, supersonic molecular beam. The chiral molecule 1-indanol is cooled to low rotational temperatures (1-2 K) and a selected rotational level in the electronic and vibrational ground state of the most abundant conformer is depleted via optical pumping on the $S_{1} \leftarrow S_{0}$ transition. Further downstream, three consecutive microwave pulses with mutually perpendicular polarizations and with a well-defined duration and phase are applied. The population in the originally depleted rotational level is subsequently monitored via laser induced fluorescence (LIF) detection. This scheme enables a quantitative comparison of experiment and theory for the transfer efficiency in what is the simplest ESST triangle for any chiral molecule, that is, the one involving the absolute ground state level, $\ket{J_{K_{a}K_{c}}} = \ket{0_{00}}$. Moreover, this scheme improves the enantiomer enrichment by over an order of magnitude compared to previous works. Starting with a racemic mixture, a straightforward extension of this scheme allows to create a molecular beam with an enantiomer-pure rotational level, holding great prospects for future spectroscopic and scattering studies. 
\end{abstract}

\maketitle

%\tableofcontents

%\section{\label{sec:level1}First-level heading:\protect\\ The line
%break was forced \lowercase{via} \textbackslash\textbackslash}

The two non-superimposable mirror images of a chiral molecule are referred to as enantiomers. Their structures cannot be transformed into each other by pure translation or rotation. Almost all physical properties of enantiomers are identical – which makes them intrinsically challenging to separate. The enantiomeric molecular pairs have identical energy levels except for a theoretically predicted tiny energy difference $(\sim10^{-11}\,\textrm{J/mol})$ induced by parity violating weak interactions \cite{letokhov1975difference,Quack2008,Darqui2010}. Yet, their interactions with other chiral objects – chiral molecules and chiral light – are dramatically different. Therefore, special care must be taken when addressing molecular chirality in biological environments since the functionality of biomolecules is dominated by the handedness of surrounding molecules \cite{polavarapu2018chiral}. 

Since Pasteur first described optical activity, various techniques such as optical rotation, circular dichroism, vibrational circular dichroism, and Raman optical activity \cite{Muller2007,he2011determination,nafie1976,stephens1985} have been developed to determine enantiomeric excess (EE) and absolute configurations of chiral molecules. Those techniques require high-density samples, but still give rise to weak signals because they rely on the inherently weak interaction of the molecules with the magnetic field of the radiation. Over the past decade, instrumental developments led to new methods which rely solely on electric dipole-allowed transitions, including Coulomb explosion \cite{Pitzer2013}, photoelectron circular dichroism \cite{powis2000photoelectron,nahon2006determination}, and microwave three-wave mixing \cite{Patterson2013a,Patterson2013b}. All these techniques can be applied in the gas phase and enable exploration of chiral molecules without interference from any surrounding physical medium. %In addition, it is possible to gain insight on the interaction of the molecules with solvents by adding these in a controlled manner \cite{LeBarbuDebus2006}. Since chiral molecules retain their handedness throughout physical processes, measurements of a sample’s enantiomeric excess carried out in the gas phase hold for the sample in solution or solid phase.

Beyond chiral analysis, chiral separation is a fundamental goal of chiral studies. In the past two decades, several optical schemes to achieve quantum-controlled chiral separation and purification have been suggested \cite{Krl2001,kral2003two,li2008dynamic}. The chiral sensitivity of these methods is due to the fact that a $C_{1}$ chiral molecule necessarily has three non-zero electric dipole moment components whose scalar triple products $(\boldsymbol{\mu}_{a}\cdot(\boldsymbol{\mu}_{b}\times\boldsymbol{\mu}_{c}))$ are opposite in sign for the two members of the enantiomeric pair \cite{Patterson2013a,Patterson2013b}. Therefore, it is possible to gain insight into the chirality of a sample regardless of the initial orientation of the molecules. These methods employ cyclic transitions in a three-level system where the interference between two different paths can be used to control the transfer of population in an enantiomer-specific fashion. 

Recently, quantum-controlled chiral separation was experimentally demonstrated using a method called enantiomer-specific state transfer (ESST) \cite{eibenberger2017enantiomer,Perez2017,Perez2018}. ESST utilizes three rotational states and three mutually orthogonally polarized microwave (MW) pulses which address all the transitions between the states \cite{Leibscher2019}. By using three resonant pulses with controlled phases, it is possible to create interfering pathways to a given target state. The model of a 3-level cyclic transition gives a qualitative understanding of ESST. However, to complete this model, $M_{J}$ degeneracy has to be taken into account. While the spatial degeneracy in ESST has been considered theoretically \cite{Lehmann2018,leibscher2020complete}, quantitative comparison between experiment and theory is still lacking. For the quantitative analysis of ESST, optimal control of the pulse sequence and pulse length of the microwave excitations is necessary. So far, ESST has been experimentally studied with sequences of partially overlapping MW pulses which hampers characterizing the effects of individual MW pulses on the interrogated molecules. In addition, poor enantiomeric enrichment makes the quantitative analysis practically difficult. The two main factors that limit the size of the population transfer are the thermal population of the three rotational energy levels participating in ESST and the $M_{J}$ degeneracy of the involved rotational states \cite{Lehmann2018,leibscher2020complete}.

\begin{figure}
\includegraphics[width=8.6cm]{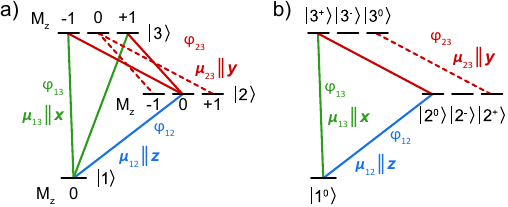}
\caption{\label{fig:1} (a) Excitation scheme of the ESST experiment for the simplest triad of rotational levels, using the z-axis for M-quantization. For polarization directions as indicated, the allowed transitions between the levels $\ket{i}$ and $\ket{j}$ together with the electric dipole moments $\mu_{ij}$ and phases $\varphi_{ij}$ are drawn in. Solid lines indicate transitions that participate in ESST, dashed lines those that do not. (b) Redefined excitation scheme, obtained by changing the basis set, resulting exclusively in two-level systems.}
\end{figure}

The most rigorous comparison between theory and experiment can be attained by studying the simplest triad which consists of the ground state $\ket{0_{00}}$ and two rotational levels with $J=1$ \cite{Ye2018}. Therefore, the simplest triad consists of either one of the following combinations: ($\ket{0_{00}}$, $\ket{1_{01}}$, $\ket{1_{10}}$), ($\ket{0_{00}}$, $\ket{1_{11}}$, $\ket{1_{10}}$) or ($\ket{0_{00}}$, $\ket{1_{01}}$, $\ket{1_{11}}$) in the $\ket{J_{{K_{a}}{K_{c}}}}$ notation. The excitation scheme used for ESST in this case is shown in Fig. 1(a) where we refer to the levels as $\ket{1}$, $\ket{2}$ and $\ket{3}$. The wavefunctions associated with these levels, $\ket{\Psi_{1}}$,$\ket{\Psi_{2}}$ and $\ket{\Psi_{3}}$, are characterized by the projection of the total angular momentum on the $z$-axis. In this triad, three resonant MW pulses are applied in the following order: $\ket{2} \xleftarrow{\pi/2} \ket{1} \xrightarrow{\pi} \ket{3} \xrightarrow{\pi/2} \ket{2}$. These three MW fields are linearly polarized and mutually orthogonal to each other. Here, the $\pi/2$ pulse creates a 50:50 superposition between two levels whereas the $\pi$ pulse inverts the population of two states. Polarization directions and non-zero dipole moment components of the MW fields are indicated next to each of the transitions in Fig. 1(a). The polarization of the first $\pi/2$ pulse defines the quantization axis ($z$-direction), and it allows $\Delta{M_z} = 0$ transitions (blue solid line). The subsequent $\pi$ and $\pi/2$ pulses are polarized along $x$- and $y$-axes and thus the selection rules are given by $\Delta{M_z}=\pm1$. Therefore, the possible transitions driven by the second MW pulse are $\ket{\Psi_{1},M_{z}=0}\rightarrow\ket{\Psi_{3},M_{z}=\pm1}$ (green solid lines). The third MW pulse connects two states with $J=1$ where four transitions are involved. Two of the transitions $\ket{\Psi_{3},M_{z}=\pm1}\rightarrow\ket{\Psi_{2},M_{z}=0}$ (red solid lines) participate in ESST, while the other two $\ket{\Psi_{3},M_{z}=0}\rightarrow\ket{\Psi_{2},M_{z}=\pm1}$ (red dashed lines) do not. 

\begin{figure*}
\includegraphics[]{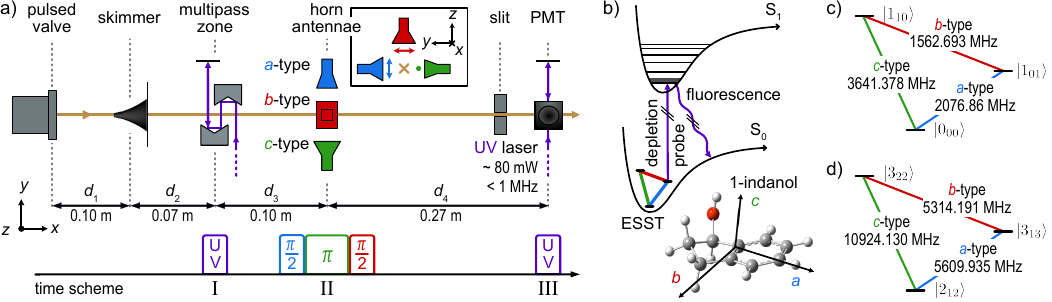}% Here is how to import EPS art
\caption{\label{fig:2} (a) Scheme of the experimental setup (not to scale). Jet-cooled 1-indanol is injected into the vacuum chamber through a pulsed valve. The molecules travel through a multipass zone where they interact with the UV depletion laser (Step I). Further downstream, three MW pulses with mutually orthogonal polarizations are applied for ESST (Step II). The configuration of these three horn antennae with their polarization directions is shown in the inset. The molecules are interrogated by the same UV laser in the detection region where the total laser induced fluorescence (LIF) intensity is measured with a photomultiplier tube (PMT) (Step III). (b) Scheme of the electronic excitation and emission processes. The most stable conformer of 1-indanol is depicted together with the inertial axis system. (c) One of the simplest triads of rotational levels of 1-indanol. Frequencies and types of each MW field are marked. (d) Energy level scheme of the second triad used, consisting of levels with $J=$ 2 and 3. Corresponding MW frequencies and types are also depicted.}
\end{figure*}

One can change the basis set and redefine the wavefunctions as follows:
\begin{subequations}
\begin{eqnarray}
\ket{\Psi_{i}^{0}}&=&\ket{\Psi_{i},M_{z}=0} \label{eq:theoryz}
\\
\ket{\Psi_{i}^{\pm}}&=&\frac{1}{\sqrt{2}}\ket{\Psi_{i},M_{z}=-1}\pm\frac{1}{\sqrt{2}}\ket{\Psi_{i},M_{z}=+1} \label{eq:theoryxy}
\end{eqnarray}
\end{subequations}
with $i = 1$, $2$ or $3$ for Eq.(\ref{eq:theoryz}) and $i = 2$ or $3$ for Eq.(\ref{eq:theoryxy}). The redefined scheme is shown in Fig. 1(b) where levels can be described as $\ket{i^{0}}$ for $i = 1$, $2$ or $3$ and $\ket{i^{\pm}}$ for $i = 2$ or $3$. It can be seen that: (i) the triangle $\ket{1^{0}}$, $\ket{2^{0}}$ and $\ket{3^{+}}$ participates in the ESST process; (ii) levels $\ket{2^{+}}$ and $\ket{3^{0}}$ constitute a closed two level system that only interacts with the last $\pi/2$ pulse; (iii) levels $\ket{2^{-}}$ and $\ket{3^{-}}$ are dark states and do not interact with the applied field.  

If relaxation processes are neglected, we can treat the ESST process as a series of two level transitions \cite{varshalovich1988}. The wavefunction of the two level system formed by $\ket{1^{0}}$ and $\ket{2^{0}}$, for instance, can be described as $a^{0}_{1}(t)\ket{\Psi^{0}_{1}} + a^{0}_{2}(t)\ket{\Psi^{0}_{2}}$, where $a^{0}_{1}(t)$ and $a^{0}_{2}(t)$ are the time-dependent probability amplitudes. The Bloch equations for resonant excitation are \cite{varshalovich1988}:
\begin{subequations}
\begin{eqnarray}
    i\hbar \dv{a^{0}_{1}(t)}{t}& =& -\frac{\Omega_{12}}{2}e^{i\varphi_{12}} a^{0}_{2}(t) \label{eq:bloch1}
    \\
     i\hbar \dv{a^{0}_{2}(t)}{t}& =& -\frac{{\Omega}  ^{\ast}_{12}}{2} e^{-i\varphi_{12}} a^{0}_{1}(t) \label{eq:bloch2}
\end{eqnarray}
\end{subequations}
where $\Omega_{12}=\mu_{12}\,\mathcal{E}_{12}/\hbar$ and $\varphi_{12}$ are the Rabi frequency and phase of the electric field driving the transition, respectively. Here, $\mu_{12}$ is the transition dipole moment and $\mathcal{E}_{12}$ is the electric field amplitude. The general solution after time $t$, for constant $\Omega_{12}$ and $\varphi_{12}$ is 
\begin{subequations}
\begin{eqnarray}
    a^{0}_{1}(t)&=& a^{0}_{1}(0)\cos \frac{\Theta(t)}{2}+i\frac{\mu_{12}}{|\mu_{12}|}e^{i\varphi_{12}} a^{0}_{2}(0)\sin\frac{\Theta(t)}{2}
    \\
      a^{0}_{2}(t)&=& i\frac{\mu^{\ast}_{12}}{|\mu_{12}|}e^{-i\varphi_{12}} a^{0}_{1}(0)\sin\frac{\Theta(t)}{2}+a^{0}_{2}(0)\cos \frac{\Theta(t)}{2}\label{eq:sol}
\end{eqnarray}
\end{subequations}
where $\Theta(t)=|\Omega_{12}|\,t/\hbar$. In the simplest case where the two upper levels $\ket{2}$ and $\ket{3}$ are empty and only the ground level $\ket{1}$ is populated, taken as $a^{0}_{1}(0) = 1$, the solution for the Bloch equations after the first $\pi/2$ pulse are: 
\begin{subequations}
\begin{eqnarray}
    a^{0}_1&=&\frac{1}{\sqrt{2}}
    \\
    a^{0}_2&=&\frac{i\mu^\ast_{12}}{\sqrt{2}|\mu_{12}|}e^{-i\varphi_{12}}
\end{eqnarray}
\end{subequations}
After the $\pi$ pulse, population inversion occurs: 
\begin{subequations}
\begin{eqnarray}
    a^{0}_1&=&0
    \\
a^{+}_3&=&\frac{i\mu^\ast_{13}}{\sqrt{2}|\mu_{13}|}e^{-i\varphi_{13}}
\end{eqnarray}
\end{subequations}
Finally, after the last $\pi/2$ pulse, the solutions are:
\begin{subequations}
	\begin{eqnarray}
	a^{0}_2&=&i\frac{\mu_{12}^\ast}{2|\mu_{12}|}e^{-i\varphi_{12}}(1\mp i e^{i (\varphi_{12} + \varphi_{23} - \varphi_{13})})
	\\
	a^{+}_3&=&i\frac{\mu_{13}^\ast}{2|\mu_{13}|}e^{-i\varphi_{13}}(1\pm i e^{i (\varphi_{12} + \varphi_{23} - \varphi_{13})})
	\end{eqnarray}\label{eq:finalRabi}
\end{subequations}
where the alternating signs for the interference term have to be used for either of the enantiomers. As a result, ESST will be achieved both in levels $\ket{2}$ and $\ket{3}$ whose populations will be given by $n_i=|a^{-}_{i}|^{2}+|a^{+}_{i}|^{2}+|a^{0}_{i}|^{2}$ for $i = 2, 3$.

When there is thermal population in level $\ket{3}$, the closed two level transition $\ket{3^{0}}\rightarrow\ket{2^{+}}$ transfers part of the population from level $\ket{3}$ to level $\ket{2}$ with the last $\pi/2$ pulse. The thermal population in level $\ket{3}$ can be expressed as $\mathcal{F}_{3}(T_\text{rot})$ times the population in level $\ket{1}$, where $T_\text{rot}$ is the rotational temperature. The total population in level $\ket{2}$ relative to the original population in level $\ket{1}$ at the end of the ESST process is then given by:
\begin{equation}
    n_2=\frac{1}{2}(1\pm\sin(\varphi_{12}+\varphi_{23}-\varphi_{13}))+\frac{1}{6}\mathcal{F}_{3}(T_\text{rot}) \label{eq:population}
\end{equation}
where $\pm$ has to be used for either enantiomer. The closed two level transition is thus seen to produce a phase-independent background.

Here we perform a quantitative study of ESST using 1-indanol. The molecules are cooled in a supersonic expansion using Ar as carrier gas $(T_\text{rot} \sim 2\,\textrm{K})$. We choose the $\ket{0_{00}}$, $\ket{1_{01}}$, and $\ket{1_{10}}$ rotational states as target triad (Fig. 2(c)). The thermal populations of all the $M_{J}$ components of these three levels are almost equal because their energy differences are small $(< 0.2\,\textrm{K})$. To increase the transfer efficiency of ESST, we developed a new spectroscopic scheme consisting out of three essential Steps (I-III) as illustrated in Fig. 2(a) and outlined below.

In Step I, a continuous-wave UV laser with an output power up to $80\,\textrm{mW}$ and a bandwidth of $< 1\,\textrm{MHz}$, is used to selectively deplete the population of the target rotational level $\ket{1_{01}}$ using the $\ket{2_{02}}\leftarrow\ket{1_{01}}$ R-branch line of the origin band of the $S_{1}\leftarrow S_{0}$ electronic transition \cite{hernandez2021high}. After UV excitation, the molecules predominantly radiate back to higher vibrational levels and to other rotational levels in the $S_{0}$ state \cite{LeBarbuDebus2006}. In this way, we optically pump and deplete all three $M_{J}$ components of the $\ket{1_{01}}$ level. We reached up to $95\%$ depletion in our experiment; for details, see the Supplemental Material.

Step II of the experiment consists of sequentially exciting three electric dipole-allowed rotational transitions ($a$-type, $b$-type, and $c$-type) with mutually orthogonal polarizations in the following manner: $\ket{1_{01}} \xleftarrow{\pi/2} \ket{0_{00}} \xrightarrow{\pi} \ket{1_{10}} \xrightarrow{\pi/2} \ket{1_{01}}$. To determine the optimal pulse duration of each MW radiation field, the relevant Rabi oscillation curves are measured prior to the ESST measurement (see the Supplemental Material). Typically, pulse lengths required for a $\pi$ pulse of each MW field are around $2\,\mu\textrm{s}$ and the ESST process is completed in less than $6\,\mu\textrm{s}$. During this short time interval, decoherence will not take place. 

In Step III, laser induced fluorescence (LIF) detection is employed to probe the population in the target rotational level $\ket{1_{01}}$. A small fraction of the power ($10\%$) of the narrow–band UV laser is used to selectively excite the same UV transition $S_{1}(2_{02})\leftarrow S_{0}(1_{01})$, probing all three $M_{J}$ levels in the target state. 

\begin{figure}
\includegraphics[width=8.6cm]{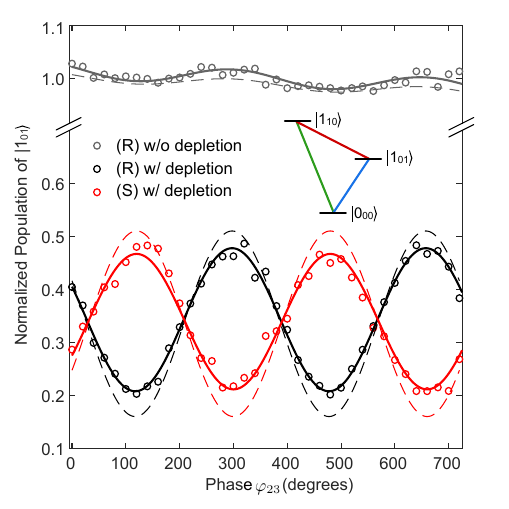}
\caption{\label{fig:3} The population of the target state $\ket{1_{01}}$ as a function of the MW field phase $\varphi_{23}$ normalized to its original thermal population. ESST results with depletion for (\textit{R})- and (\textit{S})-enantiomers are marked in black and red, respectively. Predicted ESST curves are calculated using Eq.(\ref{eq:population}) and shown with dashed lines for each case. The ESST measurement is also performed without depletion as depicted in gray. The slight overall signal decrease with increasing phase corresponds to about $3\%$ signal decrease during the measurement (about 15 minutes) and is included in the fitting.}
\end{figure} 

We perform separate measurements using enantiopure commercial samples of (\textit{R})- and (\textit{S})-1-indanol with a stated $99\%$ purity. In order to map enantiomer selectivity on the ESST signal, relative phases of the applied MW fields have to be varied as expressed in Eq.(\ref{eq:population}). For this, we keep the phases of the first two MW fields, $\varphi_{12}$ and $\varphi_{13}$, fixed and change the phase $\varphi_{23}$ in steps of 20 degrees from 0 to 720 degrees in our measurement. In Fig. 3, normalized ESST results are shown for the (\textit{R})-enantiomer in black and the (\textit{S})-enantiomer in red (details on the measurement of the normalized ESST signal are given in the Supplemental Material). As expected, the phase of the two curves differs by 180 degrees. From the best sine fits, the experimental amplitudes and offsets are determined to be 0.135(6) and 0.343(4) for the (\textit{R})-enantiomer (black solid line) and 0.128(9) and 0.340(6) for the (\textit{S})-enantiomer (red solid line), respectively. These parameters are commonly used to determine the maximum enantiomer enrichment ($\epsilon $) as $\epsilon = \textrm{amplitude/offset}$. According to this definition, we achieved 39$\%$ and 38$\%$ enantiomer enrichment for the (\textit{R})- and the (\textit{S})-enantiomers, respectively. The same ESST measurement for the (\textit{R})-enantiomer is also performed without depletion as depicted in gray in Fig. 3. The comparison of both measurements on the same $y$-scale shows the significant improvement due to the depletion process.

%Thus, the enantiomeric enrichment is just $2.9\%$, over ten times smaller than that obtained with the depletion step. This result clearly shows significance of the depletion process for the optimal ESST measurement.The best-fit experimental curve (gray solid line) has an amplitude of 0.0148 and an offset of 1.014.
The dashed lines in Fig. 3 show calculated normalized ESST curves for each case. We remove the population only from the level $\ket{1_{01}}$, whereas the level $\ket{1_{10}}$ retains its Boltzmann population. As discussed before, the $M_{z}$ = 0 level of $\ket{1_{10}}$ introduces background signal due to the last $\pi$/2 pulse that connects the $\ket{1_{01}}$ and $\ket{1_{10}}$ \textit{via} a $b$-type transition. In this case, the population of the target state $\ket{1_{01}}$ is given by Eq.(\ref{eq:population}). Note that, in Eq.(\ref{eq:population}), the population is normalized to the original population in the $\ket{0_{00}}$ level, whereas the simulation shown here is normalized to the population in the $\ket{1_{01}}$ level, assuming a Boltzmann distribution with $T_\text{rot} = 2\,\textrm{K}$. The predicted amplitude of the depleted ESST signal is 0.175 with an offset of 0.336. Comparison of the experiment with simulation shows that, for both enantiomers, the amplitude of the experimental value is $\sim20\%$ smaller than expected while the offset agrees very well. The discrepancies between experiment and simulation are most likely caused by the imperfect orthogonality between the three microwave fields. The distance of the horn antennae to the molecular beam axis is about 10 cm, which is on the same order as the wavelength of the MW radiation. We are therefore in the near field limit, probably causing the imperfect orthogonality of the polarizations of the three MW fields. Moreover, reflections of the MW radiation in the vacuum chamber cannot be completely avoided, also contributing to polarization imperfections. In addition to these effects, the stated $\sim 1\%$ impurity in our sample will contribute to the observed difference. 

%
%This is consistent with the fact that the offset will increase with the number of molecules present in $\ket{1_{01}}$ that do not participate in the population exchange, which thereby necessarily also leads to a smaller amplitude of population transfer.
\begin{figure}
\includegraphics[width=8.6cm]{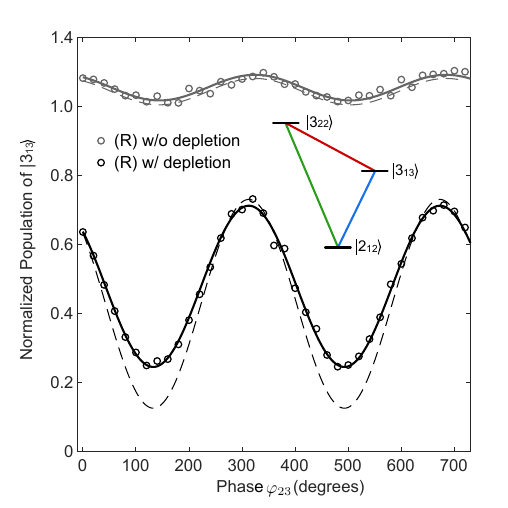}
\caption{\label{fig:4} Population of the target state $\ket{3_{13}}$ as a function of the MW field phase $\varphi_{23}$ normalized to its original thermal population. ESST results for the (\textit{R})-enantiomer with and without depletion are marked in black and gray, respectively. Predicted ESST curves are simulated and shown with dashed lines for each case.}
\end{figure}

Another triad of transitions involving the rotational energy levels, $\ket{3_{13}}$, $\ket{2_{12}}$, and $\ket{3_{22}}$ (Fig. 2(d)), is chosen for the ESST measurement. This triad involves higher $M_{J}$ degeneracy and higher MW frequencies due to the higher $J$ levels. Three resonant MW pulses are applied in the following way: $\ket{3_{13}} \xleftarrow{\pi/2} \ket{2_{12}} \xrightarrow{\pi} \ket{3_{22}} \xrightarrow{\pi/2} \ket{3_{13}}$. For this triad, the optimum pulse durations are determined from the first maximum in the Rabi oscillation curves, providing "effective" $\pi$ and $\pi/2$ pulse durations (see Supplemental Material). We measure the ESST signal by varying $\varphi_{23}$ in steps of 20 degrees from 0 to 720 degrees while the other two phases are kept fixed. In Fig. 4, normalized ESST results are shown for the (\textit{R})-enantiomer with and without depleting the target state $\ket{3_{13}}$ in black and gray, respectively. The simulated results are shown with dashed lines for each case. We achieved $\epsilon = 0.49$ for the measurement with depletion, where the simulations predict $\epsilon = 0.70$ (amplitude: 0.302; offset: 0.428). The ESST curves in this triad show an increased contrast relative to Fig. 3 because the energy separation between the $J$-levels is larger and because a larger fraction of all the involved $M_{J}$ levels can participate in the ESST process.

In summary, we quantitatively studied the ESST method by targeting the simplest triad of rotational transitions that is available for any chiral molecule. We observed $\sim20\%$ lower amplitude of the ESST signal than theoretically expected which we attribute to the imperfect orthogonality of the polarizations of the MW fields. Depleting a target rotational state prior to the ESST process enables significant enhancement in enantiomer enrichment. The enantiomeric enrichment achieved in our measurement is $49\%$ which is more than an order of magnitude larger than the largest value achieved to date, which, according to the definition used here, was $\sim3\%$ \cite{Perez2017}. When removing thermal populations of the two upper levels in the simplest triad, perfect ESST efficiency can be achieved, and experiments are underway to demonstrate this. Starting with a racemic mixture, this scheme allows to create a molecular beam with an enantiomer-pure rotational level, holding great prospects for future spectroscopic and scattering studies. 

\begin{acknowledgments}
We thank Marco De Pas, Henrik Haak, Uwe Hoppe, Sebastian Kray, Johannes Seifert and Stefan Truppe as well as the teams of the mechanical and electronics workshop of the Fritz-Haber-Institut for excellent technical and laser support.  
\end{acknowledgments}

\nocite{*}
%\bibliography{ESST} % Produces the bibliography via BibTeX.
%

\end{document}